\documentclass[%
 twocolumn,
 amsmath, amssymb,
 aps,
 pre,
 longbibliography
]{revtex4-1}

\usepackage{graphicx}
\usepackage{dcolumn}
\usepackage{bm}
\usepackage{multirow}
\usepackage{diagbox}
\usepackage[english]{babel}
\usepackage[pdfpagelabels,plainpages=false,bookmarks=true,colorlinks,linkcolor=red,urlcolor=blue,citecolor=blue]{hyperref}

\renewcommand{\vec}[1]{{\bf #1}}
\newcommand{\braket}[1]{\langle #1  \rangle}
\newcommand{\bbraket}[2]{\langle\langle #1, #2  \rangle\rangle}
\newcommand{\ccdot}{\,\cdot\,}

\newcommand{\nn}{\text{nn}}
\newcommand{\nnn}{\text{nnn}}
\newcommand{\nnnn}{\text{nnnn}}
\renewcommand{\min}{\text{min}}

\newcommand{\truncate}{\text{truncate}}

\begin{document}

\title{Determination of the Critical Manifold Tangent Space and Curvature with Monte Carlo Renormalization Group}

\author{Yantao Wu$^1$ and Roberto Car$^{1, 2}$}

\affiliation{
$^1$The Department of Physics, Princeton University\\
$^2$The Department of Chemistry, Princeton University \\
}

\date{\today}
\begin{abstract}
We show that the critical manifold of a statistical mechanical system in the vicinity of a critical point is locally accessible through correlation functions at that point. 
A practical numerical method is presented to determine the tangent space  and the curvature to the critical manifold with Variational Monte Carlo Renormalization Group. 
Because of the use of a variational bias potential of the coarse-grained variables, critical slowing down is greatly alleviated in the Monte Carlo simulation.  
In addition, this method is free of truncation error. 
We study the isotropic Ising model on square and cubic lattices, the anisotropic Ising model and the tricritical Ising model on square lattices to illustrate the method. 
\end{abstract}

\pacs{Valid PACS appear here}
\maketitle
\section{Introduction}
The introduction of renormalization group (RG) theory in statistical physics \cite{xxx} has greatly deepened our understanding of phase transitions.  
Our understanding of RG, however, is far from complete. 
The actual implementation of the RG procedure remains a highly nontrivial task.  
The critical manifold of a lattice model is defined as the set of coupling constants for which the long range physics of the system is described by a unique underlying scale-invariant field theory. 
However, the same lattice model may admit different critical behaviors described by different field theories, upon changing the coupling constants. 
This is the case, for instance, in the tricritical Ising model to be discussed later. 
Thus, the critical manifold is always defined with respect to the field theory underlying the lattice model. 
It could be defined in any space of coupling constants associated with a finite number of coupling terms, with co-dimension in that space equal to the number of relevant operators of the system.     
General RG theory requires that the RG flow should go into a unique fixed-point Hamiltonian, if the starting point of the flow is on the critical manifold. 
There are various ``natural'' RG procedures where different points on a critical manifold do not go to the same critical fixed-point, the most well-known example being the decimation rule in dimension higher than one \cite{cardy}.   
By contrast, when an RG procedure satisfies this requirement, the attractive basin of the critical fixed-point is the entire critical manifold, and a computational scheme should exist, at least in principle, to identify the critical manifold. 
Whether or not this approach can be successfully pursued were a stringent test of the RG procedure under consideration.     
Conversely, the knowledge of the critical manifold provides a straightforward way to check the validity of any RG procedure: one could simply simulate the RG flow starting from two different points in the critical manifold and verify that they eventually land on the same fixed-point.
This consideration alone should be enough motivation for developing a method to compute the critical manifold. 

Another issue for which the knowledge of the critical manifold would be of interest is the study of the geometry of the coupling constant space, i.e. the parameter manifold of a classical or quantum many-body system.    
How to define a Riemannian metric in the parameter manifold has been proposed since long time for both classical \cite{riemannian_classical} and quantum systems \cite{riemannian_quantum}. 
Recently, there have been developments in understanding the significance of the geometry of the parameter manifold for both classical and quantum systems \cite{widom_line, geometric_exponent, information_geometry, geometric_tensor,it_dg_quantum}. 
One would expect knowledge of the critical manifold would fit naturally into such developments. We do not pursue further this issue here but we leave it to future research.

In this paper, we present a method to determine the tangent space and curvature of the critical manifold at the critical points of a system with Variational Monte Carlo Renormalization Group (VMCRG) \cite{vmcrg}. 
We will show that unlike the computation of the critical exponents with Monte Carlo Renormalization Group \cite{mcrg} or VMCRG, the determination of the critical manifold tangent space (CMTS) and curvature does not suffer truncation error no matter how few renormalized coupling terms are used.   
We discuss first the case where there are no marginal operators along the RG flow, and then the case where there are.  
The examples that we consider in this paper are all classical, but the method can be extended to quantum systems if a sign-free path integral representation of the quantum system would be available.  
\section{Monte Carlo Renormalization Group and the Critical Manifold}
\subsection{Coarse-graining and Renormalized Coupling Constants}
For notational simplicity, we use the terminology for classical magnetic spins on a lattice in the following discussion, although the formalism applies in general.         
Consider a statistical mechanical system in $d$ spatial dimensions with spins $\bm\sigma$ and Hamiltonian $H^{(0)}(\bm\sigma)$,
\begin{equation}
  \label{eq:hamiltonian}
  H^{(0)}(\bm\sigma) = \sum_{\beta} K^{(0)}_\beta S_\beta(\bm\sigma)
\end{equation}
where $S_\beta(\bm\sigma)$ are the coupling terms of the system, such as nearest neighbor spin products, next nearest neighbor spin products, etc., and $\vec K^{(0)} = \{K^{(0)}_\beta\}$ are the corresponding coupling constants.     
Here we call the original Hamiltonian before any RG transformation the zeroth level renormalized Hamiltonian, hence the notation $(0)$ in the superscript.  
The critical manifold is then defined in the space of $K^{(0)}_\beta$ corresponding to a finite set of couplings $S_\beta(\bm\sigma)$. 

In a real-space RG calculation, one defines coarse-grained spins $\bm\sigma'$ in the renormalized system with a conditional probability $T(\bm\sigma' |\bm\sigma)$ that effects a scale transformation with scale factor $b$. 
$T(\bm\sigma'|\bm\sigma)$ is the probability of $\bm\sigma'$ given spin configuration $\bm\sigma$ in the original system.        
The majority rule block spin in the Ising model proposed by Kadanoff \cite{block} is one example of the coarse-grained variables. 
$T(\bm\sigma'|\bm\sigma)$ can be iterated $n$ times to define the $n$th level coarse-graining $T^{(n)}(\bm\mu|\bm\sigma)$ realizing a scale transformation with scale factor $b^n$: 
\begin{equation}
  T^{(n)}(\bm\mu|\bm\sigma) = \sum_{\bm\sigma^{(n-1)}}..\sum_{\bm\sigma^{(1)}} T(\bm\mu|\bm\sigma^{(n-1)}) \cdots T(\bm\sigma^{(1)}|\bm\sigma)
\end{equation}
$T^{(n)}$ defines the $n$th level renormalized Hamiltonian $H^{(n)}(\bm\mu)$ up to a constant $g(\vec K^{(0)})$ independent of $\bm\mu$ \cite{nauenberg}:  
\begin{equation}
  \label{eq:rg}
  \begin{split}
    H^{(n)}(\bm\mu) &\equiv -\ln \sum_{\bm\sigma} T^{(n)}(\bm\mu| \bm\sigma) e^{-H^{(0)}(\bm\sigma)} + g(\vec K^{(0)})  
\\ 
&= \sum_{\alpha} K^{(n)}_\alpha S_\alpha(\bm \mu) + g(\vec K^{(0)})
\end{split}
\end{equation}
where $\{K^{(n)}_\alpha\}$ are the $n$th level renormalized coupling constants associated with the coupling terms $S_\alpha(\bm\mu)$ defined for the $n$th level coarse-grained spins. 
Modulo the constant coupling term, $T^{(n)}(\bm\mu|\bm\sigma)$ defines $H^{(n)}(\bm\mu)$ uniquely.  
$H^{(n)}$ renormalized from different starting Hamiltonians $H^{(0)}$ will generally be different.  
However, if no marginal operators appear in the RG transformation, the renormalized Hamiltonians from different initial points on the critical manifold will converge to the same critical fixed-Hamiltonian , $H^*(\bm\mu)$, as $n$ goes to infinity.    

To probe $H^*(\bm\mu)$ in a Monte Carlo (MC) simulation, one increases the iteration level $n$ and the system size $L$, until the renormalized Hamiltonian $H^{(n)}$ becomes invariant with $n$ to the desired accuracy and the $L$ dependence becomes negligible. 
It is generally impossible to determine all of the coupling constants of $H^{(n)}(\bm\mu)$ because their number increases combinatorially with the lattice size. 
In practice, one adopts some truncation scheme and approximates $H^{(n)}$ with a finite number of coupling terms $\{S_\alpha(\bm\mu)\}$ with coupling constants $K_\alpha^{(n)}$:  
\begin{equation}
  \label{eq:limit}
  H^{(n)}(\bm\mu) \approx \sum_{\alpha} K^{(n)}_\alpha S_\alpha(\bm\mu)
\end{equation}
\subsection{Critical Manifold Tangent Space in the Absence of Marginal Operators}
To compute the CMTS, let us suppose that $K^{(0)}_\beta$ and $K^{(0)}_\beta + \delta K_\beta^{(0)}$ belong to the critical manifold and apply the RG procedure starting from these two points. 
As the difference in the irrelevant directions becomes exponentially suppressed with progressively large $n$, the corresponding two renormalized Hamiltonians will tend to the same Hamiltonian $H^{(n)}$ in the absence of RG marginal operators.  
In particular, the truncated coupling constants, $K_{\alpha, \text{truncate}}^{(n)}$ and $K_{\alpha,\text{truncate}}^{(n)} + \delta K_{\alpha,\text{truncate}}^{(n)}$, renormalized respectively from $K_\beta^{(0)}$ and $K_\beta^{(0)} + \delta K_\beta^{(0)}$, will be equal within deviations exponentially small with $n$, because they are the truncation approximation for two Hamiltonians, $H^{(n)}$ and $H^{(n)} + \delta H^{(n)}$, whose difference is exponentially small in $n$.  
Thus, the spanning set of the CMTS, $\{\delta K_\beta^{(0)}\}$, satisfies the following equation for sufficiently large $n$,  
\begin{equation}
  K^{(n)}_{\alpha,\truncate} + \sum_{\beta} \frac{\partial K_{\alpha,\truncate}^{(n)}}{\partial K^{(0)}_\beta} \delta K^{(0)}_\beta =  
  K^{(n)}_{\alpha,\truncate} 
  \label{eq:cond1}
\end{equation}
for every $\alpha$. 
That is, the CMTS $\{\delta K^{(0)}_\beta\}$ is the kernel of the $n$th level RG Jacobian:   
\begin{equation}
  \label{eq:A}
  \mathcal A^{(n,0)}_{\alpha\beta} \equiv \frac{\partial K_{\alpha,\truncate}^{(n)}}{\partial K^{(0)}_\beta}
\end{equation}
for any well-defined truncation scheme. 
In the following, we will use $K_\alpha^{(n)}$ to denote the truncated coupling constants. 

As shown in \cite{vmcrg}, VMCRG provides an efficient way to compute the renormalized constants and the RG Jacobian matrix with MC under a given truncation scheme.  
It introduces a bias potential $V(\bm\mu)$ of the coarse-grained variables, expanded in a finite set of renormalized couplings $S_\alpha(\bm\mu)$ with variational parameters $J_\alpha$: 
\begin{equation}
  V_{\vec J}(\bm\mu) = \sum_\alpha J_\alpha S_\alpha(\bm\mu), 
\end{equation}
and a variational function of $\vec J = \{J_\alpha\}$: 
\begin{equation}
  \Omega(\vec J) = \ln \sum_{\bm\mu} e^{-(H^{(n)}(\bm\mu) + V_{\vec J}(\bm\mu))} + \sum_{\bm\mu} V_{\vec J}(\bm\mu) p_t(\bm\mu) 
\end{equation}
where $p_t(\bm\mu)$ is a preset target probability distribution, which will be taken as the uniform distribution in the following. 
As proved in \cite{varyfes}, $\Omega$ is convex in each $J_\beta$, and, if one excludes the constant coupling term, has a unique minimizer, $\vec J_{\min}$, which can be found with a stochastic gradient descent algorithm using the Jacobian and the Hessian of $\Omega(\vec J)$ \cite{vmcrg}:   
\begin{equation}
\label{eq:gradient}
\frac{\partial \Omega(\vec J)}{\partial J_\alpha} = - \braket{S_\alpha(\bm \mu)}_{V_{\vec J}} + \braket{S_\alpha(\bm \mu)}_{p_t}
\end{equation}
\begin{equation}
\label{eq:hessian}
\frac{\partial^2 \Omega(\vec J)}{\partial J_\alpha \partial J_\beta} = \braket{S_\alpha(\bm \mu) S_\beta(\bm \mu)}_{V_{\vec J}} - \braket{S_\alpha(\bm \mu)}_{V_{\vec J}}\braket{S_\beta(\bm \mu)}_{V_{\vec J}}
\end{equation}
Here $\braket{\cdot}_{V_{\vec J}}$ is the biased ensemble average under $V_{\vec J}$ and $\braket{\cdot}_{p_t}$ is the ensemble average under the target probability distribution $p_t$. 
The minimizer $\vec J_\min$ then satisfies the minimizing condition: for every renormalized coupling $S_\gamma(\bm\mu)$,    
\begin{equation}
  \label{eq:min_condition}
  \braket{S_\gamma(\bm\mu)}_{V_{\min}} = \braket{S_\gamma(\bm\mu)}_{p_t}
\end{equation}
If the set of the coupling terms $S_\alpha$ is complete, $V_{\min}(\bm\mu) = \sum_{\alpha} J_{\alpha,\min} S_\alpha(\bm\mu) = -H^{(n)}(\bm\mu)$, and we identify for each $\alpha$,  
\begin{equation}
  \label{eq:rc}
  K^{(n)}_\alpha = -J_{\alpha,\min}
\end{equation}
Because the set of $S_\alpha(\bm\mu)$ is not complete, a truncation error in computing $K^{(n)}_\alpha$ is incurred. 
However, because the minimizer of $\Omega$ is unique, the truncation scheme is well-defined. 
Within VMCRG, $\mathcal A^{(n,0)}_{\alpha\beta}$ can be obtained by expanding Eq. \ref{eq:min_condition} to linear order in $\delta K^{(0)}_\beta$ and $\delta K^{(n)}_\alpha$.  
The result \cite{vmcrg} is that, for a given $\beta$, every $S_\gamma(\bm\mu)$ must satisfy,  
\begin{equation}
  \label{eq:linear}
  \sum_\alpha \bbraket{S_\gamma(\bm\mu)}{S_\alpha(\bm\mu)}_{V} \frac{\partial K_\alpha^{(n)}}{\partial K_\beta^{(0)}} = \bbraket{S_\gamma(\bm\mu)}{S_\beta(\bm\sigma)}_{V} 
\end{equation}
where $\bbraket{X}{Y}_V \equiv \braket{XY}_V - \braket{X}_V\braket{Y}_V$ is the connected correlation function of the observables $X$ and $Y$ in the biased ensemble with the potential $V_\min(\bm\mu)$.   
Thus, for any $\beta$, the Jacobian matrix element $\mathcal A^{(n, 0)}_{\alpha\beta} = \frac{\partial K_\alpha^{(n)}}{\partial K_\beta^{(0)}}$, viewed as a column vector indexed by $\alpha$, can be obtained from Eq. \ref{eq:linear} by matrix inversion. 

We also note that the method described above works for any target distribution $p_t(\bm\mu)$ in VMCRG.
A different $p_t(\bm\mu)$ will result in a different bias potential $V_{\min}(\bm\mu)$ to be used in the sampling of the matrix $\mathcal A^{(n,0)}$. 
We use the uniform distribution here because then $V_{\min}(\bm\mu)$ acts to eliminate the long-range correlation in a critical system and the resultant ensemble for the sampling of $\mathcal A^{(n,0)}$ benefits from a much faster MC relaxation \cite{vmcrg}.  
However, one can impose any arbitrary bias potential of the coarse-grained variables, $V(\bm\mu)$, and adopt the corresponding biased distribution as the target distribution.    
All the steps in the above derivation follow, and the CMTS can then be computed in the biased ensemble with the arbitrary $V(\bm\mu)$. 
In particular, if one insists on using the original ensemble with no bias potential, one only needs to set the target distribution to be the original unbiased distribution, in which case $V_{\min}$ necessarily vanishes and $\mathcal A^{(n,0)}$ is sampled in the unbiased ensemble.    
\subsection{Critical Manifold Tangent Space in the Presence of Marginal Operators}
When there are marginal operators in the RG transformation, two different points on the critical manifold will converge to different fixed-point Hamiltonians.   
However, starting from any point on the critical manifold, at sufficiently large $n$, $H^{(n)}$ will be equal to $H^{(n+1)}$, and so will the truncated renormalized constants $K_\alpha^{(n)}$ be equal to $K_\alpha^{(n+1)}$.      
Now suppose that both $K^{(0)}_\beta$ and $K^{(0)}_\beta + \delta K_\beta^{(0)}$ are on the critical manifold, respectively giving rise to the truncated renormalized constants $K^{(n)}_\alpha$ and $K^{(n)}_\alpha + \delta K^{(n)}_\alpha$.
Then, the spanning set of CMTS, $\{\delta K_\beta^{(0)}\}$, instead of Eq. \ref{eq:cond1}, satisfies the following condition, 
\begin{equation}
  K^{(n)}_\alpha + \sum_{\beta} \frac{\partial K_\alpha^{(n)}}{\partial K^{(0)}_\beta} \delta K^{(0)}_\beta =  
  K^{(n + 1)}_\alpha + \sum_{\beta} \frac{\partial K_\alpha^{(n + 1)}}{\partial K^{(0)}_\beta} \delta K^{(0)}_\beta 
\end{equation}
for every $\alpha$. 
But $K_\alpha^{(n)}$ and $K^{(n+1)}_\alpha$ are already equal up to an exponentially small difference, because they are renormalized from the same point on the critical manifold.  
Thus, when marginal operators appear in the RG transformation, the CMTS is the kernel of the matrix,  
\begin{equation}
  \mathcal A^{(n+1,0)}_{\alpha\beta} - \mathcal A^{(n,0)}_{\alpha\beta}
  \label{eq:marginal_A}
\end{equation}
\subsection{The Normal Vectors to Critical Manifold Tangent Space}
Because of the spin-flip symmetry, the renormalization of the even operators and of the odd operators are decoupled in the examples we consider here, so they can be considered separately.  
In the Ising models that we discuss later, the co-dimension of the critical manifold is one, and the tangent space is thus a hyperplane and the row vectors of $\mathcal A^{(n,0)}$ or $\mathcal A^{(n+1,0)} - \mathcal A^{(n,0)}$, for systems with or without marginal operators,  are orthogonal to this hyperplane. 
This means that the row vectors of $\mathcal{A}^{(n, 0)}$ or $\mathcal A^{(n+1,0)} - \mathcal A^{(n,0)}$ are all normal vectors to the CMTS and are parallel to one another. 
Thus, the $\mathcal P$ matrix defined as 
\begin{equation}
  \mathcal P_{\alpha\beta} = \frac{\mathcal A^{(n,0)}_{\alpha\beta}}{\mathcal A^{(n,0)}_{\alpha 1}} \text{ or } \frac{\mathcal A^{(n+1,0)}_{\alpha\beta} - \mathcal A^{(n, 0)}_{\alpha\beta}}{\mathcal A^{(n+1,0)}_{\alpha 1} - \mathcal A^{(n, 0)}_{\alpha 1}}, 
\end{equation} 
that contains the normalized row vectors of $\mathcal A^{(n,0)}$ or $\mathcal A^{(n+1,0)} - \mathcal A^{(n,0)}$, should have identical rows. 

In the tricritical Ising model that we also discuss, the critical manifold in the even subspace has co-dimension two \cite{tri_mcrg}.  
In this case, we cannot expect all the rows of $\mathcal P_{\alpha\beta}$ to be equal. 
Instead, the rows should form a two-dimensional vector space to which the CMTS is orthogonal.  
This outcome can be checked, for example, by verifying that all the row vectors of $\mathcal P_{\alpha\beta}$ lie in the vector space spanned by its first two rows. 
If such consistency checks can be satisfied, it is a testament of the validity of RG theory, which predicts that a critical fixed-point Hamiltonian exists and that the co-dimension of the critical manifold has precisely the assumed value for the models considered in this paper.    

In general, the CMTS computed from different renormalized couplings will have different statistical uncertainty because the sampling noise differs for different correlation functions in an MC simulation.   
One should, thus, trust the result with the least uncertainty and use the values computed from other renormalized constants as a consistency check.  
\section{Numerical Results for CMTS}
\subsection{2D Isotropic Ising model}
Consider the isotropic Ising model on a 2D square lattice with Hamiltonian $H(\bm\sigma)$  
\begin{equation}
  H(\bm\sigma) = -K^{(0)}_{\nn}\sum_{\braket{i,j}}\sigma_i \sigma_j - K^{(0)}_{\nnn} \sum_{[i,j]}\sigma_i \sigma_j
\end{equation}
where $\braket{i,j}$ denotes the nearest neighbor pairs and $[i,j]$ the next nearest neighbor pairs.  
$K^{(0)}_\nn$ and $K^{(0)}_\nnn$ are the corresponding coupling constants. 
This model is analytically solvable when $K^{(0)}_\nnn = 0$ and is critical at the Onsager point with $K^{(0)}_\nn = 0.4407...$ \cite{onsager}. 
Four critical points are first located with VMCRG in the coupling space of $\{K^{(0)}_\nn, K^{(0)}_\nnn\}$.   
This task can be achieved by fixing $K^{(0)}_\nnn$ and varying $K^{(0)}_\nn$ while monitoring how the corresponding renormalized coupling constant $K^{(n)}_\nn$ varies with $n$, the RG iteration index.
The largest value of the original coupling constant, $K^{(0)}_{\nn,1}$, for which $K^{(n)}_\nn$ decreases with $n$, and the smallest value, 
$K^{(0)}_{\nn,2}$, for which $K^{(n)}_\nn$ increases with $n$, define the best estimate, within statistical errors, of the interval $[K^{(0)}_{\nn,1},K^{(0)}_{\nn,2}]$ of location of the critical coupling, $K^{(0)}_{\nn,c}$. 
We notice that the calculated renormalized constants are truncated and we assume here that the truncated $K^{(n)}_\nn$ increases or decreases monotonically with the exact $K^{(n)}_\nn$. 
This assumption is very natural and does not seem to be violated in the present study.  
Alternatively, the same procedure can be performed by fixing $K_\nn$ and varying $K_\nnn$. 
In the following VMCRG calculations, we use $n = 4, L = 256$, and the $b=2$ majority rule with a random pick on tie.  
We use three renormalized couplings: the nearest neighbor product $K^{(n)}_\nn$, the next nearest product $K^{(n)}_\nnn$, and the smallest plaquette $K^{(n)}_\square$.  
The model is known to have no marginal operators. 
The four critical points shown in Table \ref{table:ising_pab} all belong to the same critical phase, as they all flow into the same truncated fixed-point renormalized Hamiltonian. 
The CMTSs are determined at these critical points in a four-dimensional coupling space spanned by $K^{(0)}_\nn$, $K^{(0)}_\nnn$, $K^{(0)}_\square$, and the third nearest neighbor products, $K^{(0)}_\nnnn$.  
The $\mathcal P_{\alpha\beta}$ is shown in Table. \ref{table:ising_pab}. 
In addition, we also show the CMTS at the Onsager point, which is analytically solvable \cite{isingcmts}.   
\begin{table}[htb!]
  \setlength{\tabcolsep}{0.8em}
\centering
  \begin{tabular}{lllll} 
    \hline
    \hline
    $K^{(0)}_\nn$ & $K^{(0)}_\nnn$ & $\mathcal P_{\alpha 2}$ & $\mathcal P_{\alpha 3}$ & $\mathcal P_{\alpha 4}$\\
   \hline
    0.4407 &0& 1.4134(3) & 0.5135(3) & 1.7963(5) \\ 
           & & 1.4146(7) & 0.5134(7) & 1.799(2) \\ 
           & & 1.413(3) & 0.511(3) & 1.794(7) \\
    Exact  & & 1.4142 & 0.5139 & 1.8006 \\ 
    \hline
    0.37   &0.0509& 1.3717(4) & 0.5242(3) & 1.7664(8) \\ 
           & &1.375(1) & 0.5243(7) & 1.773(2) \\ 
           & &1.372(4) & 0.527(3) & 1.773(6) \\
    \hline
    0.228  &0.1612& 1.2529(7) & 0.5303(4) & 1.6545(8) \\ 
           & &1.254(1) & 0.5318(8) & 1.659(2) \\ 
           & &1.252(5) & 0.535(3) & 1.65(1) \\
    \hline
    0.5  &  -0.0416& 1.4441(4) & 0.5019(5) & 1.816(1) \\ 
           & &1.444(2) & 0.503(2) & 1.818(4) \\ 
           & &1.441(7) & 0.499(6) & 1.80(1) \\
    \hline
    \hline
  \end{tabular}
  \caption{$\mathcal P_{\alpha\beta}$ for the isotropic Ising model. 
  $\alpha$ indexes rows corresponding to the three renormalized constants: $\nn, \nnn,$ and $\square$. 
  The fourth row of the table at the Onsager point shows the exact values. 
  $\beta = 2,3,$ and $4$ respectively indexes the component of the normal vector to CMTS corresponding to coupling terms $\nnn, \square,$ and $\nnnn$.   
  $\beta = 1$ corresponds to the $\nn$ coupling term and $\mathcal P_{\alpha 1}$ is always 1 by definition.  
  The simulations were performed on 16 cores independently, each of which ran $3\times 10^6$ Metropolis MC sweeps.    
  The standard errors are cited as the statistical uncertainty.  
  }
\label{table:ising_pab}
\end{table}

The CMTS can also be computed in the odd coupling subspace, as we show here for the Onsager point. 
In this calculation, we take $n = 5$, $L = 256$, and again the $b=2$ majority rule for coarse-graining. 
The CMTS in a space of four odd couplings, listed in the legend of Table \ref{table:pab_odd}, is calculated from the same four renormalized couplings. 
The result is shown in Table \ref{table:pab_odd}.  
\begin{table}[htb!]
  \setlength{\tabcolsep}{0.8em}
\centering
  \begin{tabular}{lllll} 
    \hline
    \hline
    $K^{(0)}_\nn$ & $K^{(0)}_\nnn$ & $\mathcal P_{\alpha 2}$ & $\mathcal P_{\alpha 3}$ & $\mathcal P_{\alpha 4}$\\
   \hline
    0.4407 &0& 3.31248(8) & 1.65629(4) & 1.49852(6) \\ 
           & & 3.296(2) & 1.649(4) & 1.479(2) \\ 
           & & 3.315(3) & 1.658(2) & 1.503(2) \\
           & & 3.32(5) & 1.68(4) & 1.51(3) \\
    \hline
    \hline
  \end{tabular}
  \caption{$\mathcal P_{\alpha\beta}$ for the odd coupling space of the isotropic Ising model. 
  $\alpha$ indexes rows corresponding to the four renormalized odd spin products: (0, 0), (0, 0)-(0,1)-(1,0), (0, 0)-(1, 0)-(-1,0) and (0, 0)-(1,1)-(-1,-1), where the pair $(i,j)$ is the coordinate of an Ising spin. 
  The simulations were performed on 16 cores independently, each of which ran $3\times 10^6$ Metropolis MC sweeps.    
  The standard errors are cited as the statistical uncertainty.  
  }
\label{table:pab_odd}
\end{table}

\subsection{3D Istropic Ising Model}
Consider now the same model on a 3D square lattice with $K_\nnn^{(0)} = 0$, i.e. the 3D isotropic nearest neighbor Ising model. 
This model does not have an analytical solution, but is known to experience a continuous transition at $K_\nn^{(0)} = 0.22165...$ \cite{3dising}.
To compute the CMTS at this nearest neighbor critical point, we use $n = 3$, $L = 64$, and the $b=2$ marjority rule with a random pick on tie.  
The CMTS is computed in an eight-dimensional coupling space $\{K^{(0)}\}$ spanned by the nearest-neighbor and the next nearest-neighbor renormalized coupling constants, $K^{(n)}_\nn$ and $K^{(n)}_\nnn$, as shown in Table \ref{table:3D_ising}.  
\begin{table}[htb!]
  \setlength{\tabcolsep}{0.25em}
\centering
  \begin{tabular}{llllllr} 
    \hline
    \hline
    $\mathcal P_{\alpha 2}$ & $\mathcal P_{\alpha 3}$ & $\mathcal P_{\alpha 4}$ & $\mathcal P_{\alpha 5}$ & $\mathcal P_{\alpha 6}$ & $\mathcal P_{\alpha 7}$ & $\mathcal P_{\alpha 8}$\\ 
   \hline
    2.642(8) & 1.540(8)& 6.61(3) & 2.46(1) & 0.788(3) & 6.92(4) & 1.99(1) \\ 
   \hline
    2.64(2) & 1.55(2) & 6.7(1) & 2.50(2) & 0.795(3) & 7.0(1) & 1.99(2) \\ 
    \hline
    \hline
  \end{tabular}
  \caption{$\mathcal P_{\alpha\beta}$ for the 3D isotropic Ising model. 
  The two rows in the table correspond to the two different $\alpha$ which respectively index the $\nn$ and the $\nnn$ renormalized constants.  
  $\beta$ runs from 1 to 8, corresponding to the following spin products, $S^{(0)}_\beta(\bm\sigma)$: (0, 0, 0)-(1, 0, 0), (0, 0, 0)-(1, 1, 0), (0, 0, 0)-(2, 0, 0), (0, 0, 0)-(2, 1, 0), (0, 0, 0)-(1, 0, 0)-(0, 1, 0)-(0, 0, 1), (0, 0, 0)-(1, 0, 0)-(0, 1, 0)-(1, 1, 0), (0, 0, 0)-(2, 1, 1), and (0, 0, 0)-(1, 1, 1), where the triplet $(i, j, k)$ is the coordinate of an Ising spin.         
  16 independent simulations were run, each of which took $3\times 10^5$ Metropolis MC sweeps. 
  The simulations were performed at the nearest-neighbor critical point with $K_\nn = 0.22165$. 
}
\label{table:3D_ising}
\end{table}
\subsection{2D Anistropic Ising Model}
Consider then the anisotropic Ising model on a 2D square lattice with Hamiltonian $H(\bm\sigma)$
\begin{equation}
  H(\bm\sigma) = -K^{(0)}_{\nn_x} \sum_{\braket{i, j}_x} \sigma_i \sigma_j -K^{(0)}_{\nn_y} \sum_{\braket{i,j}_y} \sigma_i \sigma_j 
\end{equation}
where $\braket{i,j}_x$ and $\braket{i,j}_y$ respectively denote the nearest neighbor pairs along the horizontal and the vertical direction. 
In the space of $\{K^{(0)}_{\nn_x}, K^{(0)}_{\nn_y}\}$, the model is exactly solvable and is critical along the line \cite{isingfermion} 
\begin{equation}
  \label{eq:ani}
  \sinh(2K_{\nn_x}^{(0)})\cdot \sinh(2K_{\nn_y}^{(0)}) = 1
\end{equation}
With the $2\times 2$ majority rule, the system admits a marginal operator due to anisotropy in the RG transformation \cite{mcrg_aniising}.   
We performed VMCRG calculations on two critical points of the system with ${K^{(0)}_{\nn_y}}/{K^{(0)}_{\nn_x}} = 2, $ and $3$, with four renormalized couplings: $K^{(n)}_{\nn_x}, K^{(n)}_{\nn_y}, K^{(n)}_{\nnn}, K^{(n)}_\square$.    
The CMTS is computed in the coupling space $\{K^{(0)}_{\nn_x}, K^{(0)}_{\nn_y}, K^{(0)}_\nnn, K^{(0)}_\square, K^{(0)}_{\nnnn_x}, K^{(0)}_{\nnnn_y}\}$ using Eq. \ref{eq:marginal_A}, as shown by $\mathcal P_{\alpha\beta}$ in Table. \ref{table:ani_pab}.  

\begin{table}[htb!]
  \setlength{\tabcolsep}{0.2em}
\centering
  \begin{tabular}{llllll} 
    \hline
    \hline
    $K^{(0)}_{\nn_x}$ & $\mathcal P_{\alpha 2}$ & $\mathcal P_{\alpha 3}$ & $\mathcal P_{\alpha 4}$ & $\mathcal P_{\alpha 5}$ & $\mathcal P_{\alpha 6}$\\
   \hline
    0.304689 & 0.653(8) & 2.387(10) & 0.814(8) & 1.749(8) & 1.21(1) \\ 
             & 0.646(4) & 2.381(5) & 0.807(4) & 1.755(4) & 1.200(5) \\ 
             & 0.647(8) & 2.38(1) & 0.808(12) & 1.747(14) & 1.20(1)\\
             & 0.63(2)  & 2.37(3) &  0.78(3)  & 1.76(4) & 1.22(3)\\
    Exact    & 0.6478 \\
    \hline
0.240606 & 0.507(4) & 2.241(5) & 0.692(7) & 1.74(1) & 0.957(7) \\
  & 0.498(2) & 2.236(3) & 0.681(3) & 1.739(3) & 0.946(4) \\
  & 0.499(8) & 2.24(1)  & 0.68(1)  & 1.736(14) & 0.940(14) \\
  & 0.500(16)& 2.23(3)  & 0.67(3)  & 1.75(4)  & 0.94(2) \\
    Exact    & 0.5 \\
    \hline
    \hline
  \end{tabular}
  \caption{$\mathcal P_{\alpha\beta}$ for the 2D anisotropic Ising model. 
  $\alpha$ indexes rows corresponding to the four renormalized constants: $\nn_x, \nn_y, \nnn,$ and $\square$. 
  $\beta = 2-6$ respectively indexes the component of the normal vector to CMTS corresponding to coupling terms $\nn_y, \nnn, \square, \nnnn_x,$ and $\nnnn_y$.   
  $\beta = 1$ corresponds to the $\nn_x$ coupling term and $\mathcal P_{\alpha 1}$ is always 1 by definition.  
  }
\label{table:ani_pab}
\end{table}
\subsection{2D Tricritical Ising Model}
Finally, let us consider the 2D tricritical Ising model with the Hamiltonian 
\begin{equation}
  H(\bm\sigma) = -K^{(0)}_\nn \sum_{\braket{i,j}}\sigma_i \sigma_j - K^{(0)}_\triangle \sum_{i} \sigma_i^2 
\end{equation}
where $\sigma = \pm 1, 0$ and $\braket{i,j}$ denotes the nearest neighbor pairs.  
In the coupling space of $K^{(0)}_\nn$ and $K^{(0)}_\triangle$, the model admits a line of Ising-like continuous phase transitions, which terminates at a tricritical point. 
At the tricritical point, the underlying conformal field theory (CFT) changes from the Ising CFT with central charge $\frac{1}{2}$ to one with central charge $\frac{7}{10}$ \cite{cft}.  
Accompanying this phase transition is a change in the co-dimension of the even critical manifold, from 1 of the Ising case to 2 of the tricritical case \cite{tri_mcrg}.   
We compute the CMTS at the tricritical point, which has been determined to occur at $K_\nn^{(0)} = 1.642(8)$ and $K_\triangle^{(0)} = -3.227(1)$ both by MCRG \cite{tri_mcrg} and finite size scaling \cite{tri_fss}. 

The coupling space we consider has six couplings, listed in Table \ref{table:tri_coup}. 
\begin{table}[htb!]
  \setlength{\tabcolsep}{0.2em}
\centering
  \begin{tabular}{ll} 
    \hline
    \hline
    \hspace{5mm} & Coupling \\
    \hline
    1  & $\sigma_i^2$\\ 
    2  & $\sigma_i\sigma_j$, $i$ and $j$ nearest neighbor\\ 
    3  & $\sigma_i\sigma_j$, $i$ and $j$ next nearest neighbor\\ 
    4  & $\sigma_i\sigma_j\sigma_k \sigma_l$, $i, j, k, l$ in the smallest plaquette\\ 
    5  & $(\sigma_i\sigma_j)^2$, $i$ and $j$ nearest neighbor\\ 
    6  & $(\sigma_i\sigma_j)^2$, $i$ and $j$ next nearest neighbor\\ 
    \hline
    \hline
  \end{tabular}
  \caption{The couplings used in the computation of CMTS for the 2D tricritical Ising model.
  }
\label{table:tri_coup}
\end{table}
We use $n = 5, L = 256$ and the $b=2$ majority-rule. 
The normal vectors to the CMTS are computed using the first five renormalized couplings, as the statistical uncertainty of the sixth renormalized coupling is too large.
The result is again represented by $\mathcal P_{\alpha\beta}$ and shown in Table \ref{table:tri_pab}. 
\begin{table}[htb!]
  \setlength{\tabcolsep}{0.2em}
\centering
  \begin{tabular}{llllll} 
    \hline
    \hline
    $\alpha$\hspace{5mm} & $\mathcal P_{\alpha 2}$ & $\mathcal P_{\alpha 3}$ & $\mathcal P_{\alpha 4}$ & $\mathcal P_{\alpha 5}$ & $\mathcal P_{\alpha 6}$\\
   \hline
    1 & 2.085(2) & 2.100(5) & 0.928(1) & 2.079(1) & 2.073(2) \\ 
    2 & 2.200(2) & 2.271(3) & 1.046(2) & 2.190(2) & 2.232(2) \\ 
    3 & 2.171(1) & 2.2285(2) & 1.0160(5) & 2.163(1) & 2.193(1)\\
    4 & 2.214(1)  & 2.283(1) &  1.04(1)  & 2.20(1) & 2.24(1)\\
    5 & 2.038(4) & 2.03(1) & 0.873(2) & 2.03(1) & 2.00(1) \\
    \hline
    \hline
  \end{tabular}
  \caption{$\mathcal P_{\alpha\beta}$ for the 2D tricritical Ising model. 
  $\alpha$ indexes rows corresponding to the first five renormalized couplings listed in Table \ref{table:tri_coup}, which also gives the couplings for $\beta = 2-6$.  
  }
\label{table:tri_pab}
\end{table}
As can be seen, the rows of $\mathcal P$ are not equal within statistical uncertainty, indicating that the co-dimension is higher than one. 
To verify that the co-dimension is two, one can check whether the row vectors for $\alpha = 3-5$ are in the vector space spanned by the first two row vectors.   
Let $\vec u_n$ be the $n$th row vector of $\mathcal P$.  
If the hypothesis of co-dimension two were correct, one could write:
\begin{equation}
  \vec u_3 = a\vec u_1 + b\vec u_2
  \label{eq:u3}
\end{equation}
and find $a$ and $b$ from the first two components of the vectors $\vec u_1, \vec u_2$, and $\vec u_3$. 
We could then check that the remaining components of $\vec u_3$ satisfy the linear relation in Eq. \ref{eq:u3} with the so found $a$ and $b$. 
A similar check can be carried out for the vectors $\vec u_4$ and $\vec u_5$. 
The vectors $\vec u_3, \vec u_4$, and $\vec u_5$ calculated in this way are reported in Table \ref{table:tri_fit}. 
\begin{table}[htb!]
  \setlength{\tabcolsep}{0.2em}
\centering
  \begin{tabular}{llllll} 
    \hline
    \hline
    $\alpha$\hspace{5mm} & $\mathcal P_{\alpha 2} \hspace{5mm}$ & $\mathcal P_{\alpha 3}\hspace{5mm}$ & $\mathcal P_{\alpha 4}\hspace{5mm}$ & $\mathcal P_{\alpha 5}\hspace{5mm}$ & $\mathcal P_{\alpha 6}$\\
   \hline
    3 & 2.171 & 2.230 & 1.019 & 2.163 & 2.194\\
    4 & 2.214 & 2.284 & 1.047 & 2.204 & 2.245\\
    5 & 2.038 & 2.026 & 0.872 & 2.033 & 2.004\\
    \hline
    \hline
  \end{tabular}
  \caption{
  $a\vec u_1 + b\vec u_2$ computed from Table \ref{table:tri_pab} for $\alpha = 3-5$ and $\beta = 2-6$.  
  }
\label{table:tri_fit}
\end{table}
As we can see, the $\mathcal P_{\alpha\beta}$ for $\alpha=3-5$ and $\beta=2-6$ in Table \ref{table:tri_fit} are equal within statistical uncertainty to the corresponding elements in Table \ref{table:tri_pab}, consistent with a co-dimension equal to two at the tricritical point.     
\section{Curvature of the Critical Manifold}
Next, we compute the curvature of the critical manifold, using the isotropic Ising model as an example. 
For a change $\{\delta K^{(0)}_\beta\}$ in the original coupling constants, we expand the corresponding change in the renormalized constants to quadratic order:   
\begin{equation}
  \label{eq:2nd_expansion}
  \delta K^{(n)}_\alpha = \sum_{\beta} \mathcal A^{(n, 0)}_{\alpha\beta} \delta K^{(0)}_\beta + \frac{1}{2} \sum_{\beta\eta} \mathcal B^{(n,0)}_{\alpha \beta\eta} \delta K^{(0)}_\beta \delta K^{(0)}_\eta
\end{equation}
where $\mathcal A^{(n,0)}_{\alpha\beta}$ and $\mathcal B^{(n, 0)}_{\alpha\beta\eta}$ can be determined by substituting Eq. \ref{eq:2nd_expansion} in Eq. \ref{eq:min_condition} and enforcing equality to second order in $\delta K^{(0)}_\alpha$. 
$\mathcal A^{(n, 0)}_{\alpha\beta}$ is already given in Eq. \ref{eq:linear}. 
The result for $\mathcal B$ is that for given $\beta$ and $\eta$, for every $\gamma$, one requires 
\begin{equation}
  \label{eq:B}
  \begin{split}
    \sum_\alpha & \bbraket{S_\gamma(\bm\mu)}{S_\alpha(\bm\mu)}_V \mathcal B^{(n,0)}_{\alpha\beta\eta} = \bbraket{S_\gamma(\bm\mu)}{S_\beta(\bm\sigma)S_\eta(\bm\sigma)}_V \\
  &+ \sum_{\alpha\nu} \mathcal A_{\alpha\beta}\mathcal A_{\nu\eta} \bbraket{S_\gamma(\bm\mu)}{S_\alpha(\bm\mu)S_\nu(\bm\mu) }_V \\
  &- 2\sum_{\alpha} \mathcal A_{\alpha\eta} \bbraket{S_\gamma(\bm\mu)}{S_\beta(\bm\sigma) S_\alpha(\bm\mu)}_V
\end{split}
\end{equation}
where the connected correlation functions are again sampled in the biased ensemble $\braket{\cdot}_V$. 
Note that $\mathcal B_{\alpha\beta\eta}$ given above is not symmetric in $\beta$ and $\eta$. 
In order for it to be interpreted as a second-order derivative, it needs to be symmetrized:  
\begin{equation}
  \label{eq:second_d}
  \frac{\partial^2 K^{(n)}_\alpha}{\partial K^{(0)}_\beta \partial K^{(0)}_\eta} = \frac{1}{2} \left(\mathcal B^{(n,0)}_{\alpha\beta\eta} + \mathcal B^{(n,0)}_{\alpha\eta\beta}\right)
\end{equation}
In the coupling space of any pair $\beta$ and $\eta$:  $\{K^{(0)}_\beta, K^{(0)}_\eta\}$, the critical manifold of the 2D isotropic Ising model is a curve, and the curvature $\kappa_{\beta\eta}$ of the critical curve can be computed from the curvature formula \cite{implicitcurvature} of the implicit curve \begin{equation}
  K^{(n)}_\alpha(K^{(0)}_\beta, K^{(0)}_\eta) =  \text{constant}
\end{equation}
with the second-order derivatives given in Eq. \ref{eq:second_d}.       
Again, this curvature is determined separately by each renormalized constant $\alpha$.   
The result is given Table \ref{table:curvature}.  
\begin{table}[htb!]
  \setlength{\tabcolsep}{0.8em}
\centering
  \begin{tabular}{lllll} 
    \hline
    \hline
    $K^{(0)}_\nn$ & \diagbox{$\beta$}{$\eta$} & $\nnn$ & $\square$ & \nnnn \\
    \hline
    0.4407 & $\nn$ &   0.143(8) & 0.27(2) & 0.21(2) \\
    &  $\nnn$ &   & 0.38(2) & 0.341(8) \\
    &  $\square$  & &  & 0.20(2) \\ 
    \multicolumn{2}{l}{Exact (\nn, \nnn)} & 0.148 & & \\ 
    \hline
    0.37 & $\nn$ & 0.18(1) & 0.23(1) & 0.30(3) \\
    &  $\nnn$ &   & 0.35(2) & 0.32(2) \\
    &  $\square$  & &  & 0.18(3) \\ 
    \hline
    0.228 & $\nn$ & 0.35(2) & 0.27(3) & 0.49(3) \\
    &  $\nnn$ &   & 0.35(4) & 0.29(2) \\
    &  $\square$  & &  & 0.20(4) \\ 
    \hline
    \hline
  \end{tabular}
  \caption{$\kappa_{\beta\eta}$ at the same three critical points as in Table \ref{table:ising_pab}, calculated from $\partial^2 K^{(n)}_\nn/\partial K^{(0)}_{\beta}\partial K^{(0)}_\eta$. The exact curvature for $\beta=\nn$ and $\eta=\nnn$ at the Onsager point is also shown \cite{isingcmts}.}
\label{table:curvature}
\end{table}
Here we only quote the result calculated from the nearest neighbor renormalized constants $K_\alpha^{(n)}$, $\alpha = \nn$. 
The curvature computed from other renormalized constants have statistical uncertainty much larger than the ones in Table \ref{table:curvature}. 

The difficulty in sampling the curvature, or generally any higher-order derivatives, compared to the tangent space, can be seen from Eq. \ref{eq:B}. 
Note that on the left side of Eq. \ref{eq:B}, the connected correlation function $\bbraket{S_\gamma}{S_\alpha}$ is of order $N$, where $N$ is the system size, but each of the terms on the right side is of order $N^2$.   
Thus, a delicate and exact cancellation of terms of order $N^2$ must happen between the terms on the right hand side of Eq. \ref{eq:B} to give a final result only of order $N$.   
The variance due to the terms on the right hand side, however, will accumulate and give an uncertainty typical for $O(N^3)$ quantities as each $S_\alpha$ is of order $N$.  
(For the CMTS, the connected correlation functions of interest are also of order $N$, but the statistical uncertainties are those typical of $O(N^2)$ quantities, as seen in Eq. \ref{eq:linear}.) 
In general, as an $m$-th order derivative of the critical manifold is computed, the connected correlation functions of interest will always be of order $N$, but the correlation functions that need to sampled will be of order $N^{m+1}$, giving an exceedingly large variance.  
Thus, although in principle arbitrarily high order information about the critical manifold is available by expanding Eq. \ref{eq:min_condition}, in practice only low-order knowledge on the critical manifold can be obtained with small statistical uncertainty from a simulation near a single critical point.      

\section{Conclusion}
We have presented an MC procedure to obtain the local geometrical information on the critical manifold in the vicinity of a given critical point. 
The procedure is in essence a projector Monte Carlo method that is based on the fact that the irrelevant operators in a system decay exponentially fast along an RG trajectory. 
Because of such decay, the truncated RG Jacobian matrix, $\mathcal A^{(n,0)}$, acquires a structure that is asymptotically clearer and clearer as $n$ increases, i.e. its kernel emerges with co-dimension equal to the number of relevant operators of the system.       
This structure is quite robust. 
On the one hand, it is immune from the truncation of the renormalized Hamiltonian. 
On the other hand, it does not depend on what biased potential of the coarse-grained variables is applied to the system.     

From the perspective of connected correlation functions between the orignal spins $\bm\sigma$ and the coarse-grained spins $\bm\mu$, the aforementioned structure means the following. 
Given any bias potential $V(\bm\mu)$ at any critical point, each local observable $S_\beta$ of $\bm\sigma$ can be viewed as a linear functional $\bbraket{\ccdot}{S_\beta(\bm\sigma)}$ on the space of the local observables of $\bm\mu$:   
\begin{equation}
  \bbraket{\ccdot}{S_\beta(\bm\sigma)}: S_\gamma(\bm\mu) \mapsto \bbraket{S_\gamma(\bm\mu)}{S_\beta(\bm\sigma}_V
\end{equation}
The presence of the CMTS implies that many distinct linear functionals are linearly dependent. 
In fact, by Eq. \ref{eq:linear}, for any $\{\delta K_\beta^{(0)}\}$ in the CMTS,       
\begin{equation}
  \label{eq:condition}
  \sum_\beta \bbraket{\ccdot}{S_\beta(\bm\sigma)} \delta K^{(0)}_\beta = 0
\end{equation}
This poses an infinite number of conditions which the coarse-graining procedure has to satisfy to generate a proper RG structure.    
The majority-rule coarse-graining considered in our examples seems to do very well in satisfying these conditions. 
But a question still remains. 
Are the conditions satisfied exactly or just approximately but so closely that any violation is overshadowed by the statistical uncertainty?     
In the latter case, which coarse-graining procedure, preferably with a finite number of parameters, can satisfy all the conditions in Eq. \ref{eq:condition}?    
In the former case, what is the profound reason why all these conditions can be satisfied simultaneously?   
\begin{acknowledgments}
All the codes used in this project were written in C\texttt{++}, and will be available upon request. We acknowledge support from DOE Award DE-SC0017865.  
\end{acknowledgments}
\bibliographystyle{apsrev}

\end{document}